\documentclass[twocolumn,letter]{jpsj3}

\usepackage{color}
\usepackage{graphicx}
\usepackage{dcolumn}
\usepackage{bm}

\begin{document}

\title{Suppression of Nonmagnetic Insulating State by Application of Pressure in Mineral Tetrahedrite Cu$_{12}$Sb$_{4}$S$_{13}$}

\author{Shunsaku~Kitagawa$^{1,2}$\thanks{E-mail address: shunsaku@science.okayama-u.ac.jp}, Taishi~Sekiya$^{1}$, Shingo~Araki$^{1,2}$, Tatsuo~C.~Kobayashi$^{1,2}$, Kenji~Ishida$^{3}$, Takashi~Kambe$^{1,2}$, Takumi~Kimura$^{1}$, Naoki~Nishimoto$^{1}$, Kazutaka~Kudo$^{1,2}$, and Minoru~Nohara$^{1,2}$}
\inst{$^1$Department of Physics, Okayama University, Okayama 700-8530, Japan \\
$^2$Research Center of New Functional Materials for Energy Production, Storage and Transport,
Okayama University, Okayama 700-8530, Japan \\
$^3$Department of Physics, Kyoto University, Kyoto 606-8502, Japan}

\date{\today}

\newcommand{\red}[1]{\textcolor{red}{#1}}

\abst{
The mineral tetrahedrite Cu$_{12}$Sb$_{4}$S$_{13}$ exhibits a first-order metal--insulator transition (MIT) at $T_{\rm MI}$ = 85~K and ambient pressure.
We measured the $^{63}$Cu-NMR at ambient pressure and the resistivity and magnetic susceptibility at high pressures.
$^{63}$Cu-NMR results indicate a nonmagnetic insulating ground state in this compound.
The MIT is monotonically suppressed by pressure and disappears at $\sim1.0$~GPa.
Two other anomalies are observed in the resistivity measurements, and the pressure -- temperature phase diagram of Cu$_{12}$Sb$_{4}$S$_{13}$ is constructed.
}

\abovecaptionskip=-5pt
\belowcaptionskip=-10pt
\setlength{\textheight}{660pt}

\maketitle

Mixed-valent transition-metal (TM) compounds with a fractional number of electrons per TM site often exhibit a metal--insulator transition (MIT) associated with charge ordering.
The insulating ground state is either nonmagnetic or magnetically ordered depending on the compound.
Typical compounds with nonmagnetic ground states include Magneli phase Ti$_{4}$O$_{7}$ with Ti$^{3.5+}$ (3$d^{0.5}$)~\cite{S.Lakkis_PRB_1976}, oxide spinel AlV$_{2}$O$_{4}$ with V$^{2.5+}$ (3$d^{2.5}$)~\cite{Y.Horibe_PRL_2006}, and thiospinel CuIr$_{2}$S$_{4}$ with Ir$^{3.5+}$ (5$d^{5.5}$)~\cite{P.G.Radaelli_Nature_2002}.
For instance, CuIr$_{2}$S$_{4}$ experiences an MIT at 230~K associated with the charge ordering of the mixed-valent Ir ion.
The formation of isomorphic octamers of Ir$^{3+}_{8}$S$_{24}$ and Ir$^{4+}_{8}$S$_{24}$ with the dimerization of Ir$^{4+}$ (5$d^{5}$, $S$ = 1/2) pairs results in a spin-singlet ground state~\cite{P.G.Radaelli_Nature_2002,K.Takubo_PRL_2005}.
On the other hand, vanadium bronze $\beta$-Na$_{0.33}$V$_{2}$O$_{5}$ with (1/6)V$^{4+}$ + (5/6)V$^{5+}$ (3$d^{1/6}$)~\cite{T.Yamauchi_PRL_2002}, chromium hollandite K$_{2}$Cr$_{8}$O$_{16}$ with Cr$^{3.75+}$ (3$d^{2.25}$)~\cite{K.Hasegawa_PRL_2009}, and magnetite Fe$_{3}$O$_{4}$ with Fe$^{2.5+}$ (3$d^{5.5}$) for a B sublattice~\cite{E.J.W.Verwey_Nature_1939,M.S.Senn_Nature_2012} exhibit magnetically ordered ground states.
Interestingly, superconductivity appears in $\beta$-Na$_{0.33}$V$_{2}$O$_{5}$~\cite{T.Yamauchi_PRL_2002} and CuIr$_{2}$S$_{4}$~\cite{N.Matsumoto_PRB_1999,G.Cao_PRB_2001} when charge ordering is suppressed by applying hydrostatic pressure and chemical doping, respectively.

\begin{figure}[!b]
\vspace*{10pt}
\begin{center}
\includegraphics[width=8cm,clip]{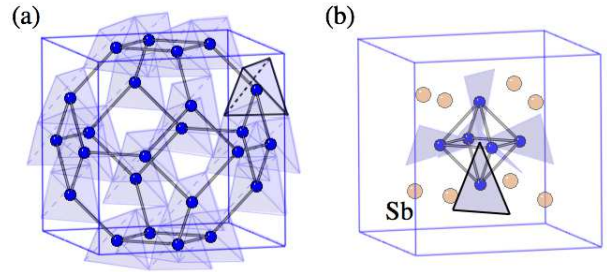}
\end{center}
\caption{(Color online) Crystal structure of tetrahedrite Cu$_{12}$Sb$_{4}$S$_{13}$: (a) Cu(1) and (b) Cu(2) substructures. A Cu(2) tetrahedron at the body-centered site is shown in (b). 
Cu(2) tetrahedra at the corners are not shown for clarity.
}
\label{Fig.1}
\end{figure}

The mineral tetrahedrite Cu$_{12}$Sb$_{4}$S$_{13}$ is another example of a mixed-valent TM compound and has been attracting considerable interest because of its enhanced thermoelectric properties~\cite{K.Suekuni_APE_2012,K.Suekuni_JAP_2013,X.Lu_JEM_2013,X.Lu_AEM_2013,J.Heo_CM_2014}.
Apart from its thermoelectricity, Cu$_{12}$Sb$_{4}$S$_{13}$ exhibits intriguing physical properties such as an MIT and a simultaneous sudden decrease in the magnetic susceptibility $\chi$ at 85~K~\cite{K.Suekuni_APE_2012,F.D.Benedetto_PCM_2005}, which is reminiscent of thiospinel CuIr$_{2}$S$_{4}$~\cite{N.Matsumoto_PRB_1999,G.Cao_PRB_2001}.
However, understanding this transition is not intuitive because of the small fraction of magnetic Cu$^{2+}$ (3$d^{9}$, $S$ = 1/2) in Cu$^{2+}_{2}$Cu$^{+}_{10}$Sb$^{3+}_{4}$S$^{2-}_{13}$ with respect to the formal valence.
Tetrahedrite Cu$_{12}$Sb$_{4}$S$_{13}$ crystallizes in a body-centered cubic structure with the space group $I\bar{4}3m$ (No. 217)~\cite{B.J.Wuensch_Science_1963,B.J.Wuensch_ZKr_1964}.
The structure consists of Cu(1)S$_{4}$ tetrahedra and Cu(2)S$_{3}$ planar triangles, where Cu(1) and Cu(2) occupy crystallographic 12$d$ and 12$e$ sites, respectively.
The edge-sharing Cu(1)S$_{4}$ tetrahedra form a cage structure of truncated octahedra of Cu$_{24}$, while Cu(2)S$_{3}$ triangles in the Cu(1) cage form Cu$_{6}$ octahedra, as shown in Fig.~\ref{Fig.1}.
Cu(1) is either divalent (Cu$^{2+}$, 3$d^{9}$) or monovalent (Cu$^{+}$, 3$d^{10}$), while Cu(2) is monovalent according to coordination chemistry~\cite{F.D.Benedetto_CM_2002}.
This odd fractional number of electrons per Cu(1) site, i.e., (1/3)Cu$^{2+}$ + (2/3)Cu$^{+}$ (3$d^{10-1/3}$), raises the question of whether charge ordering and magnetic ordering are involved in the MIT of Cu$_{12}$Sb$_{4}$S$_{13}$.

In this paper, we report on the $^{63}$Cu-NMR of the MIT of tetrahedrite Cu$_{12}$Sb$_{4}$S$_{13}$ at ambient pressure and the effects of pressure on the MIT via resistivity $\rho$ and $\chi$ measurements.
$^{63}$Cu-NMR spectra revealed that the ground state at ambient pressure is nonmagnetic; this suggests the formation of a spin singlet state at the MIT of 85~K.
Furthermore, we observed that the MIT is suppressed by the application of hydrostatic pressure; the nonmagnetic insulating ground state switches with a metallic ground state at $\sim1.0$~GPa.
Above 1.26~GPa, we discovered another anomaly that is characterized by a peak in the electrical resistivity $\rho$.
This peak temperature monotonically increases from $\simeq$ 85 to 170~K at a pressure of up to 4~GPa.
No superconductivity was observed down to 0.1~K in this metallic phase at high pressures.

Polycrystalline samples of Cu$_{12}$Sb$_4$S$_{13}$ were synthesized through the heating of a mixture of CuS, Cu$_2$S, and Sb$_2$S$_3$ powders.
Stoichiometric amounts of the powders were mixed, pulverized, and sealed in an evacuated quartz tube.
The ampules were heated to 900 $^\circ$C at a rate of 100 $^\circ$C/h and then cooled to 600 $^\circ$C at a rate of 2 $^\circ$C/h followed by furnace cooling.
The resulting samples were dark gray with metallic luster.
They were characterized by powder X-ray diffraction analysis with a Rigaku RINT-TTR III X-ray diffractometer using Cu $K_\alpha$ radiation and identified to be Cu$_{12}$Sb$_4$S$_{13}$ with a small amount of CuSbS$_2$.

Electrical resistivity was measured with a standard four-probe method.
The magnetic susceptibility was measured with a superconducting quantum interference device magnetometer (MPMS; Quantum Design) under a constant magnetic field $\mu_{0} H$ = 1~T.
An indenter-type pressure cell and Daphne 7474 as the pressure-transmitting medium were utilized to measure the resistivity at high pressures~\cite{T.C.Kobayashi_RSI_2007,K.Murata_RSI_2008}.
A $^{3}$He-$^{4}$He dilution refrigerator was used in the resistivity measurement down to 0.1~K.
A piston-cylinder-type pressure cell and Daphne 7373 were utilized for the magnetic susceptibility measurement at high pressures.
The applied pressure was estimated from the $T_{\rm c}$ of the lead manometer~\cite{A.Eiling_JPFMP_1981,B.Bireckoven_JPESI_1988}.

A conventional spin-echo technique was utilized for the subsequent NMR/nuclear quadrupole resonance (NQR) measurement.
We used silver as the NMR coil to avoid background signals from a Cu coil.
The nuclear spin $I$ and the nuclear gyromagnetic ratio $\gamma$ of $^{63}$Cu are 3/2 and 11.285 MHz/T, respectively.
When $I \ge 1$, the nucleus has an electric quadrupole moment $Q$ as well as a magnetic dipole moment; thus, the degeneracy of the nuclear-energy levels is lifted by both Zeeman and electric quadrupole interactions.
The Hamiltonian can be described as
\begin{align}
\mathcal{H} &= \mathcal{H}_{\rm Z} + \mathcal{H}_{\rm Q} \notag \\
            &= -\gamma \hslash (1 + K)I \cdot H \notag \\
& + \frac{\hslash \nu_{zz}}{6}\left\{3I_{z}^{2}-\bm{I}^{2} + \frac{1}{2}\eta (I_{+}^{2} + I_{-}^{2}) \right\},
\end{align}
where $\mathcal{H}_{\rm Z}$ and $\mathcal{H}_{\rm Q}$ are the Zeeman and electric quadrupole Hamiltonians, $K$ is the Knight shift, $H$ is the external field, $\nu_{zz} (\propto V_{zz})$ is the quadrupole frequency along the principal axis of the electric field gradient (EFG), and $\eta (= |V_{xx}-V_{yy}|/V_{zz})$ is the asymmetric parameter of the EFG.
Here, $V_{ij} \equiv \partial ^2V/\partial x_{i}\partial x_{j}$  ($V$ is the electrostatic potential; $x_{i}, x_{j} = x, y, z$) is the EFG tensor.

\begin{figure}[!t]
\vspace*{-0pt}
\begin{center}
\includegraphics[width=8.5cm,clip]{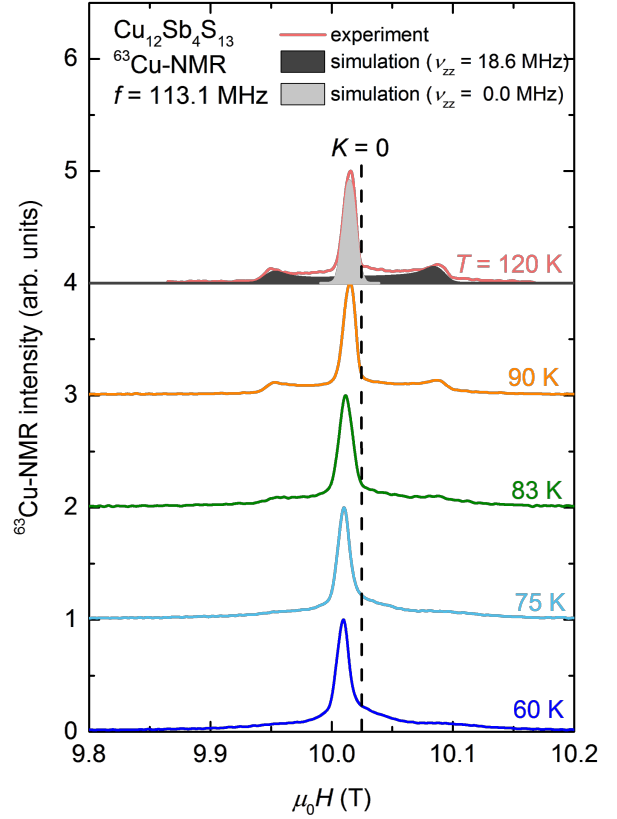}
\end{center}
\vspace*{-0pt}
\caption{(Color online) Field-swept $^{63}$Cu NMR spectra at different temperatures in Cu$_{12}$Sb$_{4}$S$_{13}$.
Simulations are shown with NMR spectra at 120~K.
The obtained quadrupolar parameters $\nu_{zz}$ were $\sim$~0 and 18.6~MHz.
}
\label{Fig.2}
\end{figure}
\begin{figure}[!t]
\vspace*{-0pt}
\begin{center}
\includegraphics[width=8.5cm,clip]{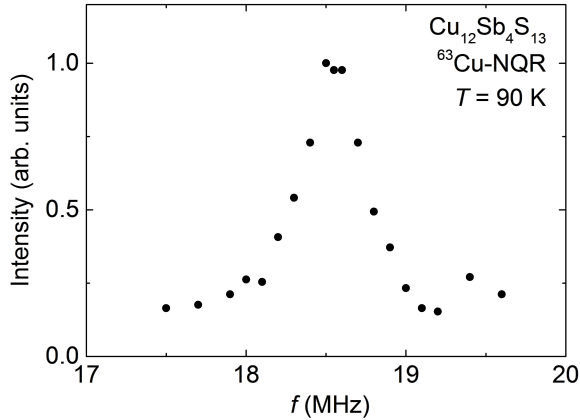}
\end{center}
\vspace*{-0pt}
\caption{Frequency-swept $^{63}$Cu NQR spectra at 90~K for Cu$_{12}$Sb$_{4}$S$_{13}$.
}
\label{Fig.3}
\end{figure}

With regard to the results, we first focus on the physical properties at ambient pressure.
Figure~\ref{Fig.2} shows the temperature dependence of field-swept $^{63}$Cu-NMR spectra.
Three clear peaks were observed above the MIT temperature $T_{\rm MI}$ of 85 K.
These peaks were due to the superposition of two NMR spectra arising from the  $^{63}$Cu(1) and $^{63}$Cu(2) sites:
We can fit the spectra by two central transition lines ($I_{z} = 1/2 \leftrightarrow -1/2$) to the $^{63}$Cu~($I$ = 3/2).
The NMR spectra were perturbed by the two different quadrupolar parameters $\nu_{zz}$ of $\sim$ 0 and 18.6~MHz.
This was also confirmed in the NQR measurement as shown in Fig.~\ref{Fig.3}.
We could not observe any satellite transition lines  ($I_{z} = \pm 3/2 \leftrightarrow \pm 1/2$) probably owing to the large distribution of EFG.
Thus, we did not determine $\eta$.
The ratio of the signal intensity of two simulated central transition lines is almost 1:1, which corresponds to that of Cu(1) and Cu(2) site abundances expected from the crystal structure.
We considered which Cu site corresponds to each quadrupolar parameter.
Because Cu(1) is located at the center of a nearly regular tetrahedron, as shown in Fig.~\ref{Fig.1}(a), $\nu_{zz}$ should be relatively small.
Then, $\nu_{zz} \sim 0$~MHz should correspond to the Cu(1) site.
On the other hand, Cu(2) is located on the planar triangle of S.
Therefore, the $\nu_{zz}$ of the Cu(2) site should be large, corresponding to 18.6~MHz.
CuS (covellite) consists of CuS$_4$ tetrahedra and CuS$_3$ planar triangles, and similar $^{63}$Cu-NMR spectra were reported~\cite{Y.Itoh_JPSJ_1996}.

\begin{figure}[!tb]
\vspace*{10pt}
\begin{center}
\includegraphics[width=8.7cm,clip]{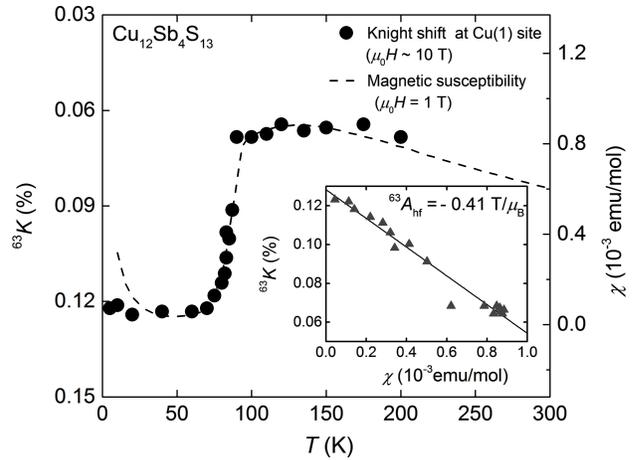}
\end{center}
\caption{Temperature dependences of Knight shift $K$ at Cu(1) site at 10~T and of $\chi$ at 1~T for Cu$_{12}$Sb$_{4}$S$_{13}$.
The inset shows the $K$--$\chi$ plot above 40~K.
The solid line portrays the fitting line.
}
\label{Fig.4}
\end{figure}

The $^{63}$Cu-NMR measurement at ambient pressure revealed that the ground state below $T_{\rm MI}$ is not a magnetic insulator but a nonmagnetic one.
The temperature dependence of the Knight shift $K$, which reflects the local susceptibility at the nuclear site, was determined from the peak field of the NMR spectra with $\nu_{zz} \sim 0$~MHz and is shown in Fig.~\ref{Fig.4}.
The temperature dependence of $\chi$ is also plotted and found to be consistent with previous reports~\cite{K.Suekuni_APE_2012,F.D.Benedetto_PCM_2005}.
Since a low-temperature upturn is not observed in the Knight shift, it would come from impurity spins.
$K$ was well proportional to $\chi$ above 40~K, although the applied magnetic field differed between $K$ and $\chi$.
The inset of Fig.~\ref{Fig.4} plots $K$ against $\chi$, which gives an estimation of the hyperfine coupling constant $A_{\rm hf}$ = $-$~0.41 T/$\mu_{\rm B}$.
This value doubles when we assume that two Cu$^{2+}$ spins/f.u., whose number is expected from charge neutrality, contribute to magnetic properties.
The negative hyperfine coupling constant can be explained by a dipole effect.
$K$ rapidly increased below $T_{\rm MI}$, which indicates that the spin susceptibility decreased.
More importantly, the shape of the $^{63}$Cu-NMR spectra with $\nu_{zz} \sim 0$~MHz remained almost unchanged across $T_{\rm MI}$, which indicates that an internal field does not exist at the Cu(1) site below $T_{\rm MI}$.
Note that the powder patterns of the NMR spectra should exhibit a drastic change in their shape when magnetic ordering occurs: it is well known that the spectral shape becomes nearly rectangular in a commensurate antiferromagnetic (AFM) state, and that the linewidth broadens in the incommensurate AFM state.
Our $^{63}$Cu-NMR results demonstrate that there was no AFM order and that the Cu(1) site became nonmagnetic below $T_{\rm MI}$ in Cu$_{12}$Sb$_{4}$S$_{13}$.
The signal intensity of $^{63}$Cu-NMR spectra with $\nu_{zz} \sim 0$~MHz also does not change markedlly across $T_{\rm MI}$, indicating that a signal from both Cu$^{+}$ and Cu$^{2+}$ is observed even below $T_{\rm MI}$.
In addition, NMR spectra with $\nu_{zz}$ = 18.6~MHz became broader and smeared out below $T_{\rm MI}$, which indicates that the EFG distribution at the Cu(2) site became larger.
The NQR spectra also smeared out below $T_{\rm MI}$.

\begin{figure}[!tb]
\vspace*{-0pt}
\begin{center}
\includegraphics[width=8.5cm,clip]{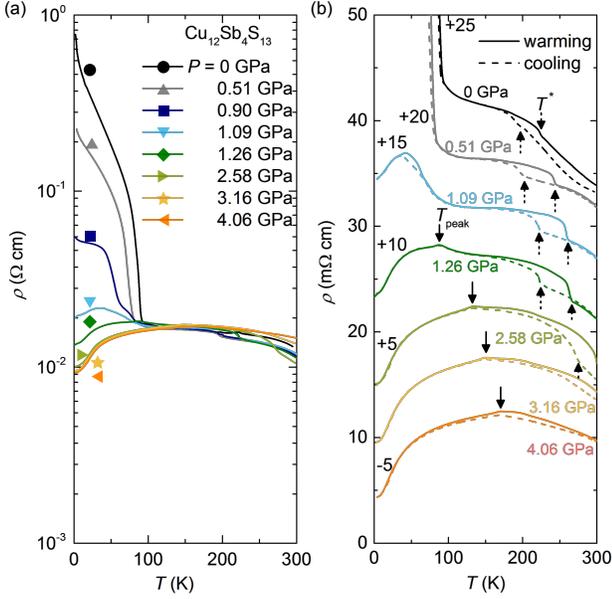}
\end{center}
\caption{(Color online) (a) Temperature dependences of $\rho$ under different pressures. 
(b) Temperature dependences of $\rho$ focused on low-$\rho$ region in Cu$_{12}$Sb$_{4}$S$_{13}$.
The solid and broken arrows indicate $T_{\rm peak}$ and $T^{*}$, respectively.}
\label{Fig.5}
\end{figure}

\begin{figure}[!tb]
\vspace*{15pt}
\begin{center}
\includegraphics[width=8.5cm,clip]{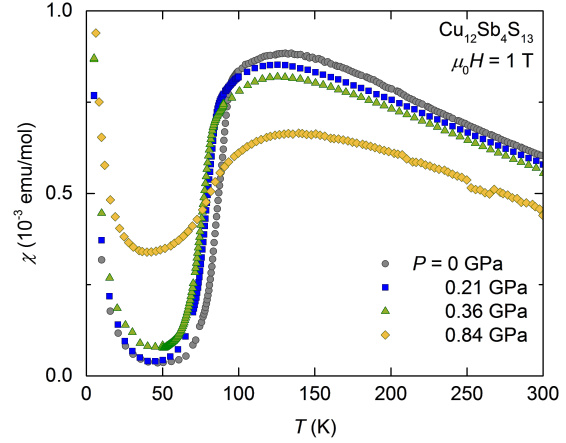}
\end{center}
\caption{(Color online) Temperature dependences of $\chi$ under different pressures in Cu$_{12}$Sb$_{4}$S$_{13}$.}
\label{Fig.6}
\end{figure}

\begin{figure}[!tb]
\vspace*{-0pt}
\begin{center}
\includegraphics[width=9.2cm,clip]{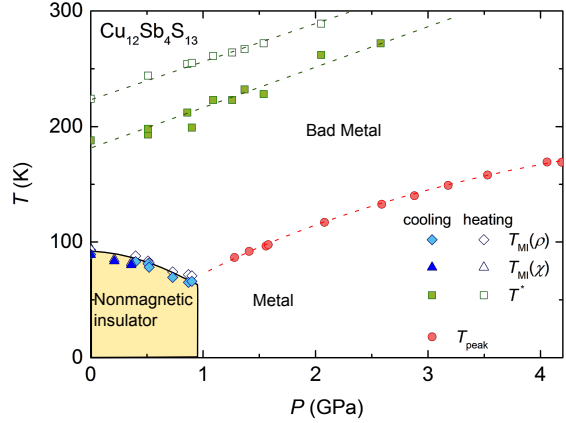}
\end{center}
\caption{(Color online) $P$--$T$ phase diagram of Cu$_{12}$Sb$_{4}$S$_{13}$.
The diamonds and triangles represent metal--insulator transition temperature $T_{\rm MI}$ estimated from $\rho$ and $\chi$, respectively.
The circles and squares represent $T_{\rm peak}$ and $T^{*}$ estimated from $\rho$, respectively. 
The broken line is a visual guide.
}
\label{Fig.7}
\end{figure}

Next, we present the pressure dependence of the MIT.
The nonmagnetic insulating ground state became metallic above $\sim$1~GPa.
Figures~\ref{Fig.5}(a) and \ref{Fig.5}(b) show the temperature dependence of $\rho$ under different pressures.
The MIT was observed at 85~K and ambient pressure.
$T_{\rm MI}$ monotonically decreased with increasing pressure from 85~K at ambient pressure to 65~K at 0.90~GPa.
Similar results are obtained by the magnetic susceptibility measurement as shown in Fig.~\ref{Fig.6}.
Above 1.26~GPa, the MIT was completely suppressed, and $\rho$ exhibited a clear peak at approximately 100~K.
Below the temperature of the peak $\rho$ value, $\rho$ monotonically decreases on cooling.
We defined this peak as $T_{\rm peak}$.
$T_{\rm peak}$ increased with pressure up to 4.19~GPa.
Similar results were reported by Umeo {\it et~al}.~\cite{K.Umeo_JPS_2013}.
The residual resistivity $\rho_0$ monotonically decreased with increasing pressure.
There was no sign of superconductivity down to 2~K at all measurement pressures and down to 0.1~K at 1.26, 1.37, and 3.16~GPa.
As shown in Fig.~\ref{Fig.5}(b), an additional anomaly with hysteresis is observed at high temperatures.
This anomaly temperature is defined as $T^{*}$.
A similar anomaly is observed at ambient pressure in a previous report~\cite{K.Suekuni_APE_2012}.
$T^{*}$ increases with increasing pressure.
$\chi$ does not show any anomaly at $T^{*}$, as shown in Fig.~\ref{Fig.6}.
The origins of $T_{\rm peak}$ and $T^{*}$ are unclear.
We summarize the pressure--temperature phase diagram of Cu$_{12}$Sb$_{4}$S$_{13}$ in Fig.~\ref{Fig.7}.

Here, we discuss possible mechanisms of the MIT with a nonmagnetic ground state in Cu$_{12}$Sb$_4$S$_{13}$.
One possibility is the charge order state at the Cu(1) site.
According to the coordination chemistry, the Cu(1) site is a mixed valent where one-third of Cu is Cu$^{2+}$ (3$d^{9}$, $S$ = 1/2) and two-thirds is Cu$^{+}$ (3$d^{10}$, $S$ = 0).
Because Cu$^{2+}$ has a $S$ = 1/2 moment, a spin singlet state can be formed owing to the dimerization of Cu$^{2+}$ ions, and a nonmagnetic insulating state results.
This is reminiscent of CuIr$_{2}$S$_{4}$, where the formation of Ir octamers with the dimerization of Ir$^{4+}$ ($5d^5$, $S$ = 1/2) pairs results in a spin-singlet ground state~\cite{P.G.Radaelli_Nature_2002,K.Takubo_PRL_2005}.
In the case of Cu$_{12}$Sb$_{4}$S$_{13}$, one-third of Cu(1) with a cage structure, as shown in Fig.~\ref{Fig.1}(a), is Cu$^{2+}$.
Realizing an appropriate charge ordering of Cu$^{2+}$ and Cu$^{+}$ with a ratio of 1:2 on the Cu(1) cage structure, so that the dimerization of adjacent Cu$^{2+}$ ions to form a spin singlet ground state is not straightforward.
The significant broadening of the Cu(2) NMR spectra with $\nu_{zz}$ = 18.6~MHz may be related to the random dimerization of Cu(1)$^{2+}$.
The random dimerization does not induce any broadening of Cu(1) NMR spectra because of the negligible $\nu_{zz}$ at the Cu(1) site.

In conclusion, we measured the temperature dependence of $^{63}$Cu-NMR spectra at ambient pressure and of $\rho$ and $\chi$ at high pressures in tetrahedrite Cu$_{12}$Sb$_{4}$S$_{13}$ in order to investigate the origin and pressure dependence of the MIT.
The $^{63}$Cu-NMR results indicated that the ground state at ambient pressure is a nonmagnetic insulating state.
$T_{\rm MI}$ decreased with increasing pressure and switched to $T_{\rm peak}$ above $\sim1.0$~GPa.
Cu$_{12}$Sb$_{4}$S$_{13}$ shows a unique MIT that was characterized as a phase transition from a paramagnetic bad metal to a nonmagnetic insulator at $T_{\rm MI}$.
In addition, this nonmagnetic insulating ground state changed to a metallic state at a relatively low applied pressure.
A novel type of charge ordering may be realized for Cu$_{12}$Sb$_{4}$S$_{13}$.

\section*{Acknowledgments}
We are grateful to H. Matsuno for discussion.
This work was partially supported by the Okayama University Cryogenic Center, and Grants-in-Aid from the Japan Society for the Promotion of Science (JSPS) (Grant Nos. 15H01047, 23244075, 25400372, and 26287082).

\end{document}